\documentclass{elsart}

 \usepackage{graphicx}

\usepackage{amssymb}

\usepackage{subfigure}

\begin{document}

\begin{frontmatter}



\title{ Networks of Mobile Elements for Biological Systems}



 \author{Kanchan Thadani\corauthref{cor1}}
 \ead{kanchant@cdac.in}
 \corauth[cor1]{}
 \address{Centre for Development of Advanced Computing \\Pune University Campus \\ Pune 411007, Maharashtra, India\\}

 \author{Ashutosh}
 \ead{ashutosh\_a@persistent.co.in}
 \address{Persistent Systems, \\3-B Bhageerath, Senapati Bapat Road \\ Pune 411008, Maharashtra, India}
 
\begin{abstract}

In this paper we present a network model to study the impact of spatial distribution of constituents, coupling between them  and diffusive processes in the context of biological situations. The model is in terms of network of mobile elements
whose internal dynamics is given by differential equations exhibiting switching and/or oscillatory behaviour. To make the model more consistent with the underlying biological phenomena we incorporate properties like growth and decay into the network. Such a model exhibits a plethora of attributes which are interesting from both the network theory perspective as well as from the point of view of bio-chemistry and biology.

We characterise this network by calculating the usual network measures like network efficiency, entropy growth, vertex degree distribution, geodesic lengths, centrality as well as fractal dimensions and generalised entropy. It is seen that the model can demonstrate the features of both scale free networks as well 
as small worlds network in different parameter domains. 
The formation of coherent spatio-temporal patterns is another feature of such networks which makes them appealing for understanding broad qualitative aspects of problems like  cell differentiation (e.g. morphogenesis) and synchronization (e.g. quorum sensing mechanisms). 

 One of the key features of any biological system is its response to external attacks. The response of the network to various attack strategies(isolated and multiple) is also studied. 

\end{abstract}


\end{frontmatter}

\section{Introduction}
\label{Intro}

Biological systems exhibit tremendous complexity, versatility and robustness in their responses to their environment. Though molecular biology has had spectacular success it is becoming clearer that simple enumeration of genes, proteins and metabolites 
is not sufficient to understand this complex behaviour. In recent time it has been shown that \cite{CellNet} to capture system level behaviour, it is instructive to search for patterns common to complex systems and networks in general. For example, even if we could calculate the physio-chemical properties of a protein from its structure, to obtain the physiological behaviour from these would be a daunting task without the knowledge about the organisation of the network in which the proteins operate. Networks have been discovered to be important at various levels in biological systems \cite{Hartwell}. These networks of interacting elements are intrinsically dynamic and describe how the system changes in space and time in response to the external stimuli, grows and reproduces, differentiates and dies. Therefore  there is a pressing need  in biology to develop an alternative language and methodology to deal with such situations.

{\it ``Perhaps a proper understanding of the complex regulatory networks making up cellular systems like the cell cycle
will require a shift from common sense thinking. We might need to move into a strange more abstract world, more
readily analyzable in terms of mathematics than our present imaginings of cells operating as a microcosm of our
everyday world" } \cite{Nurse} \\

{\it ``The best test of our understanding of cells will be to make quantitative predictions about their behaviour and test
them. This will require detailed simulations of the biochemical processes taking place within [cells]. We need to
develop simplifying, higher-level models and find general principles that will allow us to grasp and manipulate the
functions of [biochemical networks]."} \cite{Hartwell}

This modelling of processes in cellular and molecular biology is based upon the assumption that interactions between molecular components can be approximated by a network of biochemical reactions in an ideal macroscopic reactor. Then following the standard methods of chemical reaction kinetics, ordinary differential equations are obtained. This modeling approach has been applied to many systems.

However, there are a few considerations which are omitted in above approximation. Firstly, though a few studies consider the spatial aspects of the problem (e.g. \cite{Eldar} on {\it Drosophila melanogaster}), most of the time they are simply neglected. However, it is known that the cell is not a homogenous well stirred reactor, but instead is a highly compartmentalized and heterogenous environment where phenomena like crowding or channeling are present \cite{Ellis}. In such a situation, the spatial aspects may well play a key role and may need to be incorporated. Such spatial models are particularly relevant for biological networks where the concentration and growth of signalling factors decays with the distance. Thus the probability of establishing a connection with far off nodes is much smaller. Thus the nodes in such networks are more likely to be directly connected to the ``nearby nodes'' than in the far off nodes. The spatial inhomogeneity observed in real life networks is also indicative of the fact that the values in the network are not uniformly distributed but instead clustered together at various points in space. There may also be bounds and limits in spatial model which forbid or allow connections depending on a cut-off distance.

Another important consideration is that of mobility. In most of the studies the elements are considered to be fixed in space and the structure of coupling is fixed in time i.e. two elements that are initially coupled are coupled forever and they remain at the same point in space. Some models do consider time dependent coupling coefficients \cite{kaneko} though spatial positions are fixed. However, in real situations the elements move due to diffusion or chemical gradients \cite{chemotatctic} and the {\it local interactions} determine the connectivity of the element.

There is another way in which such models make restrictive assumptions in context of biological systems.
The number of nodes and degrees of freedom are fixed initially and remain fixed. However most of the biological systems show growth, reproduction and death. For example a cell multiplies and transfers its attributes to its offspring and dies. There may be more than one pathway to produce a protein and also more than one way it is degraded. The biological elements also differentiate themselves to allow flexibility with respect to function \cite{Differentiate}. This differentiation and growth leads to  degrees of freedom no longer remaining fixed for a node and diversification of the nature of the elements themselves.

Lastly, most of the models assume either a boolean function for evolution \cite{boolean} or an arbitrary  dynamical rule like dynamical maps(e.g. logistic, circle or Cellular Automata update rules) or some convenient rule like piecewise linear differential equations \cite{piecewise}. However, such a dynamical rule should have some correspondence(e.g. same kind of non-linearity, continuity properties) to the underlying biological consideration so that it can reproduce behaviour more realistically. Ideally this function should be inferred from experimental data or the detailed knowledge of underlying kinetics. However, as an approximation, at least some motif biological evolution rules which are found in a number of contexts should be used to model the broad qualitative features with a hope that they fall in the same class of systems  as the one being modelled and reproduce results with a greater degree of faithfulness to reality.

Thus in light of the above discussion, there is a need to generalise the existing network models of the biological systems. Here we attempt to relax all of the assumptions made above and study the behaviour of such models.

The paper is organised as follows: In the next section we set up the network and elaborate on its features.
We also discuss the internal dynamics of the nodes and various choices one can make. The section on analysis studies the  mathematical properties of the network and compares them with numerical results. After that we consider various strategies of attack on the network to determine its robustness. Finally we conclude with discussion and interpretation of results.

\section{Materials and methods}
\label{MaM}

In this section we set up the network as follows:

\subsection{Defining Network Behaviour}

A network consists of a set of connected nodes. Each node has an internal state and an intra-dynamics which evolves the state according to a dynamical system. The current state of the node is also dependent on the state of all its neighbours by  coupling. In the network that we set up, the number of neighbours a node is coupled to is dependent on the dynamics of the network (inter-dynamics). This is in contrast to the lattice models where the number of neighbours is fixed a-priori.  In such lattice models, one either couples to a fixed number of nearest neighbours or has a global coupling to all the other nodes in the network (mean-field models). We feel that in a number of biological contexts all the above assumptions are a little too restricted. During the course of time evolution, a given sub part of the system usually interacts with a variable number of other elements \cite{kaneko}. Furthermore the network that we set up is an evolving one, where new nodes get added and redundant nodes get removed. Therefore, the number of neighbours should be dynamically changing. 

Evolving networks have been studied in recent times in great detail \\ \cite{networks}. Typically the networks evolve by adding nodes to a random initial configuration (e.g. Erdos-Renyi network model). The nodes are usally added by using different heuristic prescriptions. In particular, {\it scale free networks} are obtained by adding new nodes based on the vertex degree distribution. These networks have been subject of much study and debate \cite{scalefree} and model many biological contexts. for example phenomena involving  food webs, biochemical interactions, protein-protein interaction maps for viruses, prokaryotes and eukaryotes C.elegans etc are described by scale free networks.  A different prescription (e.g Watt's Beta model) results in a {\it small worlds network} which is  found to be useful in modeling co-operative behavior in biological systems \cite {networks}. However, due to finite size and limited data available in most realistic situations, it is difficult to establish scale free behavior of a network unambiguously \cite{scalefree}. Secondly, the scale free nature is often inherent in the prescription to grow a network.

We take a different approach in the present work. Rather than starting from a random network, we start with a network of few elements and then grow the network by state-splitting \cite{lindandmarcus}. Such a growth law does not have a built in prescription for power law behaviour. Thus any scale free nature found is truly an emergent property rather than a consequence of a heuristic for network growth. 

Diffusion plays a major role in most of the biological processes. Thus any model which describes a process involving spatial motion or distribution of biological elements should take into account this important feature. The existing network models try to incorporate this by varying the connections between the elements dynamically \cite{kaneko}. However the assumptions that are made in these models are too restrictive in many situations.We describe these assumptions and their possible relaxations below.

\begin{enumerate}

\item  
\textbf{State is mostly considered to be discrete.} For example most commonly used network for modelling is a boolean network in which internal state of the nodes is a discrete (0 or 1) state. This kind of model is, at best, an approximation for modelling qualitative coarse grained behaviour of the system. Some of the limitations of a boolean approach are the following : firstly since the internal state of an elemnent is evolving in a discrete manner, it becomes diffficult to model situations which involve elements with different times scale being updated asynchronously. Secondly, we can not have a coupling between elements which is weighted by the continous value of the internal state. Such a coupling is desirable particularly in bio-chemical reaction networks Thirdly, as a dynamical system, the nodes can have more than one attractor i.e. more than one possible final state depending on initial values. To model such situations with boolean networks is cumbersome. On the other hands, using differential equation has advantages of preserving "quantitativity and causality" \cite{dehoon} 
In more realistic models, the nodes have a number of states possible. We assume the state variable describing the internal state to be a continuous variable thus allowing more quantitative approach.
 
\item 
\textbf{The internal state of the node does not affect the interactions in the networks.} In realistic situations this assumption is restrictive. This assumption is actually inherent in the nature of the dynamical systems models where couplings are usually considered as fixed. There are models which consider variable topologies \\ \cite{networks}. Even in such models the nodes are usually kept fixed and the connections that they make are dynamically changing. However, in most of the situations the elements move (either along a signal gradient or in a diffusive manner) in a very heterogenous medium. Most of the interactions are also dependent on the actual concentrations of active features of these elements. These active features depend on the internal state of the node and change with time. Thus the number of other nodes a given element would interact with depends on the current values of its internal state as well as the current values of internal states of the other elements. Therefore there is a need for introducing an explicit coupling between the internal states and the strength of interaction. We address this issue as follows: The nodes are allowed to move according to either a diffusive motion or a gradient or both. The interaction that takes place depends upon the number of other nodes in its ``sphere of influence'' and the internal states of the interacting elements have an effect on each other. The mobility of the element is dependent on its internal state and the region it finds itself in. Thus we have an explicit coupling between the number of neighbours that a node has and its internal state.

\item  
\textbf{The interaction range is independent of time.} However, for biological systems one can not categorise them as having a fixed interaction range but should have coupling terms which are changing with time. This is because such systems usually have a number of spatio-temporal scales with different elements interacting at different scales. This is modelled in the current system by using a coupling strength that is dependent on the distance between the interacting elements. This dependence can incorporated  either by using a threshold value or by using an interaction strength which decays with distance. Thus as an element moves around, its interaction range with other elements is dynamically changing. 

\item 
\textbf{The structure of interactions does not change with time} i.e. if two elements are coupled then they have no mechanism to decouple and two decoupled elements do not couple. We allow for such a mechanism. If an element is outside the sphere of influence its interaction diminishes and if it moves within sphere of influence of other element it gets coupled. Thus the overall network dynamics encompasses many coupling and decoupling events and overall evolution of network travels through many structures of interactions.

\item 
\textbf{Though there is a growth law (fixed by network evolution heuristic) there is usually no decay law.} As mentioned earlier elements can grow, reproduce and die in many situations of interest. The decay law, modelling the removal or death of elements in biology plays an important role in determining behaviour. We incorporate these effects as follows: the growth is modelled by state splitting of the network \cite{lindandmarcus}. Such a mechanism is natural for these contexts. There are two kinds of state splitting that are considered. In-splitting in which the incoming degree of a node is split between the node and its offspring and the outgoing degree is shared by both and the out-splitting in which outgoing degree is split and incoming connections are shared. This is different from self-organised scale free network in which nodes get all the degrees inherited. The decay is modelled as following: once a node is out of the interaction range of all other nodes, it decouples and is marked as dying. In such a state it would evolve as an independent dynamical system and reach a attractor of its evolution state and stay at that state without contributiong to the dynamics on the network. However, since other nodes are also mobile it can happen that another node with connections to the network comes within the sphere of influence of this node and get coupled to it. Then the node is ``healed'' and starts contributing to the network dynamics again. The nodes which do not heal within a specified time are considered dead and are removed from the system. Such a removal does not affect the dynamics of the rest of the system and is analogous to how biological systems clear dead and redundant parts.

\item 
\textbf{The elements are considered identical.} We incorporate diversity and differentiation of the elements in our system. Firstly, the network is made up of a number of different types of elements which are representative of various kinds of dynamics actually observed in biological networks (see next section). 
\end{enumerate}

Thus our network is described by the following: We have a set of coupled nodes. The number of nodes each element couples to is determined by its state and its sphere of influence. Every node also has a variable defining its position in space. This variable depends on the current state of the element, current position of the element, current position of its neighbours and current state of the neighbours. Thus the interaction structure of the network is dynamically dependent on the states, positions and spheres of influence of the nodes.
 
The network grows by reproduction of its nodes. To model this growth we use state splitting and to model decay we use the following rule: if two nodes are greater than a certain distance apart then they are set to be decoupled. A node which has been decoupled from every node is a ``dying''   node. However, due to dynamic interactions, such a node may still be revived (``healed''). Any dying node which does not heal for a specified period is considered ``dead'' and removed from the network.
 
  Thus the network is described by the following set of equations

Calculation of the state of a node.
\begin{equation}
{\bf X}_{j}^{t+1} = {\bf X}_{j}^{t} + (1 / N_{1_j}^{t}) \sum_{i=1}^{{N_{1_j}^{t}}} \left ( \epsilon_{in} (i) f^{[i]}({\bf X_i^t}) \right )  - (1 / N_{2_j}^{t}) \sum_{i=1}^{{N_{2_j}^{t}}} \left ( \epsilon_{out} (i) f^{[i]}({\bf X_i^t}) \right )
\label{neteq1}
\end{equation}

Calculation of the current position of a node.
\begin{equation}
 r_{j}^{t+1} = r_{j}^{t} + \sum_{i=1}^{N_{1_j}^{t}} \left ( \epsilon_{in} (i) {{\sqrt{{\bf X_i}.{\bf X_j}}} \over {| r_i^t -r_j^t|}   } \right )   + \sum_{i=1}^{N_{2_j}^{t}} \left ( \epsilon_{out} (i) {{\sqrt{{\bf X_i}.{\bf X_j}}} \over {| r_i^t -r_j^t|}   } \right )
\label{neteq2}
\end{equation}

Calculation of the number of in-neighbours and out-neighbours of a node.
\begin{equation}
N_{1_j}^{t+1} = 2 \left [ \sum_{i=1}^{N_{1_j}^t}  \theta(| r_i^t - r_j^t|) \right ] exp( - c /4)
\label{neteq3}
\end{equation}
	
\begin{equation}
  N_{2_j}^{t+1} = 2 \left [ \sum_{i=1}^{N_{2_j}^t}  \theta(| r_i^t -r_j^t|) \right ] exp( - c /4) 
\label{neteq4}
\end{equation}

where

${\bf X}_{j}^{t} \rightarrow$ The state vector for a node j at time t

$r_{j}^{t} \rightarrow$ The position of a node j at time t

$N_{1_j}^{t} \rightarrow$ The number of in neighbours of a node j at time t

$N_{2_j}^{t} \rightarrow$ The number of out neighbours of a node j at time t

$\epsilon_{in} (i) \rightarrow$ The in coupling coefficient for node i

$\epsilon_{out} (i) \rightarrow$ The out coupling coefficient for node i

$ c(R) \rightarrow$ is a coefficient depending on radius of influence R with value between 0 and 1.

\begin{eqnarray}
\theta(| r_i -r_j|) &= & 0 \hskip 0.1in if | r_i -r_j| > R \nonumber \\
                &= & 1 \hskip 0.1in otherwise
\end{eqnarray}

\subsection{Defining Node Behaviour}

Every node has an internal state and an internal dynamics. For the internal dynamics we choose some well known motifs which are frequently seen in dynamics of signalling and regulatory pathways in the cells \cite{tyson}. These motifs, which behave as switches and oscillators, serve as building blocks for the bigger, more complex networks. For example, consider a typical situation

The proteins that modulate Cdc2 activity are themselves modulated by \\ Cdc2:Cdc13, through a
set of feedback loops
\begin{enumerate}
 \item Rum1 inhibits Cdc2:Cdc13, but Cdc2:Cdc13 phosphorylates Rum1, thereby targeting Rum1 for degradation.
\item Ste9:APC labels Cdc13 for degradation, but Cdc2:Cdc13 can phosphorylate Ste9, thereby downregulating its activity and targeting it for degradation.
\item Wee1 phosphorylates and inactivates Cdc2:Cdc13, but, at
the same time, Cdc2:Cdc13 is trying to phosphorylate and
inactivate Wee1.
\item Cdc25 takes the inactivating phosphate group off PCdc2:
Cdc13, and Cdc2:Cdc13 returns the favor by phosphorylating
and thereby activating Cdc25.
\item  Slp1:APC, which also labels Cdc13 for degradation, is itself
activated by Cdc2:Cdc13 by an indirect pathway.
\end{enumerate}

The first three feedback loops are examples of mutual antagonism. Under appropriate conditions, the antagonists cannot coexist, i.e. the feedback loop works like a toggle switch. Either Cdc2:Cdc13 has the upper hand and its antagonist (Rum1 or Ste9 or Wee1) is suppressed, or vice versa. The fourth interaction is a positive feedback loop: Cdc2 and Cdc25 activate  each other in a mutually amplifying fashion. The last interaction is a time-delayed negative feedback loop, which, under appropriate conditions, can generate oscillations (as Cdc2:Cdc13 concentration rises, it turns on Slp1, which targets Cdc13 for degradation, causing Cdc2:Cdc13 concentration to fall, and Slp1 to turn off)

To study the dynamical consequences of these feedback loops, one must formulate these interactions as a precise molecular mechanism, convert the mechanism into a set of nonlinear ordinary differential equations, and study the solutions of the differential equations by numerical simulation.

We can set up a network by randomly selecting these elements and connecting them. However, for the present study we choose four elements behaving as switches and two behaving as oscillators.

 The equations for the switches and oscillators that we use are as follows
 
 {\it Perfectly Adapted Switch}
 
 Perfect adaptation means that the steady state response of the element is independent of the signal strength. Such a behaviour is typical of chemotactic systems (including human sense of smell) which initally responds to an abrupt change in interacting attractants and then settles down into a steady response \cite{chemotatctic}, \cite{lev} have used a mechanism of this sort to model phosphoinosityl signaling in slime mold cells and neutrophils. The equations representing such a system are
 
  \begin{eqnarray}
   {dX_{1} \over dt} &=& {k_{1} S - k_{2} X_{1} X_{2} } \\
   {dX_{2} \over dt} &=&  {k_{3} S - k_{4} X_{1}  } \nonumber  \\ 
    k_{1} &=& k_{2} = 2, k_{3} = k_{4} =1, S = 1.2 \nonumber 
  \label{PASwitch}
  \end{eqnarray}

In some situations, like Frog oocyte maturation in response to progesterone and Apoptosis the switching is one way. At a critical signal strength we get an irreversible transition to large response starting from small responses. The transition can be activating or inhibitory giving rise to mutual activation or inhibition
 
{\it Mutual Activation Switch}
 
The equations representing such a system are

  \begin{eqnarray}
   {dX_{1} \over dt} &=&  {k_{0} E(X_{1}) - k_{2}E(X_{1}) X_{1} } \\
    E(X_{1}) &=& GBK(k_3 X_{1}, k_{4}, J_{3}, J_{4})\nonumber  \\
    k_{0} &=& 0.4, k_{1} = 0.1, k_{2} = k_{3} =1, k_{4} = 0.2,\nonumber  \\
    J_{3} &=& J_{4} = 0.05, S = 9.0 \nonumber 
  \label{MASwitch}
  \end{eqnarray}

The Goldbeter- Koshland function \cite{gbk} is  graded and reversible. By graded we mean that the response increases continuously with signal strength. A slightly stronger signal gives a slightly stronger response. Reversible means that if we go from an initial value to a final value. Although continuous and reversible, a sigmoidal response is abrupt. Tyson \cite{tyson} compares this to buzzer or a laser pointer, to activate the response one must push hard enough on the button, and
to sustain the response one must keep pushing. When one lets up on the button, the response switches off at precisely
the same signal strength at which it switched on.
  
  However, even in the presence of such a reversible term, the above switch shows
  irreversible behaviour due to a feedback loop that is present.
  
{\it Mutual Inhibition Switch}

In this switch, if signal is decreased enough, the switch will go back to the off-state. For intermediate stimulus strengths the response of the system can be either small or large, depending on how S was changed. This switch shows hystersis and is found in the lac operon in bacteria \\ \cite{lac}, the activation of M-phase-promoting factor (MPF) in frog egg extracts \cite{frog}, the autocatalytic conversion of normal prion protein to its pathogenic form \\ \cite{prion} and budding yeast cell cycle. This behaviour has been observed in a number of experiments 
 The equations representing this system are
 
  \begin{eqnarray}
   {dX_{1} \over dt} &=&  {k_{0}+ k_{1} S  - k_{2} X_{1} - k_{2}^{'} E(X_{1})X_{1}  } \\
    E(X_{1}) &=& GBK(k_3 , k_{4} X_{1}, J_{3}, J_{4}) \nonumber  \\
    k_{0 } &=& 0.0, k_{1} = 0.5, k_{2} = 0.1, k_{3} =1, k_4 = 0.2, \nonumber  \\
    J_{3} &=& J_{4}=0.05, k_{2}^{'} = 0.5, S = 10.0 \nonumber 
  \label{MISwitch}
  \end{eqnarray}

{\it Negative FeedBack Switch}
This type of regulation, commonly employed in biosynthetic pathways, is called homeostasis. 

 The equations representing this system are

  \begin{eqnarray}
   {dX_{1} \over dt} &=&  {k_{0} E(X_{1}) - k_{2} S X_{1} } \\
    E(X_{1}) &=& GBK(k_3 , k_{4} X_{1}, J_{3}, J_{4}) \nonumber  \\
    k_{0 } &=& 1.0, k_{2} = 1.0, k_{3} =0.5, k_4 = 1.0, \nonumber  \\
    J_{3} &=& J_{4} = 0.01, S = 0.1 \nonumber 
  \label{NFSwitch}
  \end{eqnarray}

{\it Activator Inhibitor Oscillator}

The classic example of an activator-inhibitor system is cyclic AMP production in the slime mold, Dictyostelium
discoideum \cite{slime}. External cAMP binds to a surface receptor, which stimulates adenylate cyclase to produce and
excrete more cAMP. At the same time, cAMP-binding pushes the receptor into an inactive form. After cAMP falls off, the inactive form slowly recovers its ability to bind cAMP and stimulate adenylate cyclase again. This mechanism lies behind all the curious properties of the cAMP signaling system in Dictyostelium: oscillations, relay, adaptation, and wave propagation.

 The equations representing this system are
 
  \begin{eqnarray}
   {dX_{1} \over dt}&=&  {k_{0} E(X_{1}) + k_{1}S - k_{2} X_{1} -k_{2}^{'} X_{1}X_{2} }  \\
   {dX_{2} \over dt} &=&  {k_{5} X_{1} - k_{6} X_{2}  } \nonumber  \\ 
    E(X_{1}) &=& GBK(k_3 X_{1} , k_{4}, J_{3}, J_{4}) \nonumber \\
    k_{0 } &=& 4.0, k_{1} = k_{2} = k_{2}^{'} =  k_{3} =  k_4 = 1.0, k_{5} = 0.1, k_{6} =0.075, \nonumber \\
    J_{3} &=& J_{4} = 0.3, S = 0.2 \nonumber 
  \label{AIOscillator}
  \end{eqnarray}

{\it Substrate Depletion Oscillator}

 This describes MPF oscillations in frog egg extracts \cite{frog}. MPF is a dimer of a kinase sub-unit, cyclin-dependent kinase 1 (Cdk1), and a regulatory subunit, cyclin B. As cyclin B accumulates in the extract, it combines rapidly with Cdk1 (in excess). The dimer is immediately inactivated by phosphorylation of the kinase subunit X can be converted into active MPF  by a phosphatase called Cdc25. Active MPF activates Cdc25 by phosphorylating it.
 
 The equations representing this system are

  \begin{eqnarray}
   {dX_{1} \over dt}&=& { k_{1}S - (k_{0}^{'} + k_{0}E(X_{2}) X_{1} } \\
   {dX_{2} \over dt}&=& {(k_{0}^{'} + k_{0}E(X_{2}) X_{1} - k_{2} X_{1} }  \nonumber \\ 
    E(X_{1}) &=& GBK(k_3 X_{1} , k_{4}, J_{3}, J_{4})  \nonumber \\
     k_{0}^{'}&=& 0.01, k_{0} = 0.4, k_{1} = k_{2}  =  k_{3} =  1.0 \nonumber \\
     k_{4} &=& 0.3, J_{3} = J_{4} = 0.05, S = 0.3  \nonumber 
  \label{SDOscillator}
  \end{eqnarray}

where the $GBK(u,v,J,K)$ is the Goldbeter-Koshland term
\begin{equation}
 GBK(u,v,J,K) = {{2 u k} \over {v - u + v J + u K + \sqrt{{v - u + v J + u K}^2 - 4 (v-u) uK }}}  
 \end{equation}

 Thus we can model a number of situations by coupling together these motifs. In the current work, we choose a pair of any of these elements to represent the types of nodes in our network. The parameters specified are fixed at these values following \cite{tyson}. The nodes are evolved by integrating the equations using a fourth order Runge-Kutta routine for 100 iterates with time step 0.1 for the switches and 0.001 for oscillators.

 \subsection{Evolving the Network}
 We follow the algorithm outlined in flow chart of Fig. 1. The algorithm evolves each node for a fixed time, updates the number of its neighbours Eq. (\ref{neteq3}), Eq. (\ref{neteq4}), update the current state by coupling it to the neighbours by Eq. (\ref{neteq1}), obtain the current position of the node by Eq. (\ref{neteq2}) and finally update the network by splitting the in and out degree. This consitutes one iteration of the network dynamics. We set the simulation as follows: we choose two of the above six types of nodes and set them up according to an initial golden mean network. The coupling coefficient $\epsilon$ for this particular simulation is considered same for every node, though in principle the equations allow us to have a different value of  $\epsilon$ for every node. The offspring of each node is of the same as parent. We choose all possible pairs of the node to do the simulation.

 \section{Characterisation and Analysis}
 
 To understand the behaviour of a network a number of characterizing quantities have been introduced in recent times \\ \cite{networks}. It is important to choose the set of characterisers which reflect the essential features of the network with respect to the problems of interest. We classify the characterisers as the following : {\it Topological characterisers } which contain the information about the underlying topological structure of the connections in the network,{\it Geometric characterisers} , the growth or {\it Entropic characterisers} and characterisers for the dynamics. These characterisers are listed in Table 1 for three representative sets of parameters.
 
 \subsection{Topological characterisers}
 
 {\sc Degree Distributions}
 \vskip 0.1in
 The degree is an important characteristic of a node. Based on the degree of the nodes, it is possible to construct measurements for the network. One of the simplest is the maximum degree which is simply the maximum connections that any node in the network has. The degree distribution, $ P(k)$ which expresses the fraction of nodes in a network with degree k contains more information.
 One may be interested in finding out if there is a correlation between the
 degrees of different nodes. Such correlations were found to have an important
 role in many network structural and dynamical properties \cite{networks}. The most common choice is to find correlations between two nodes connected by a link. This
 correlation can be expressed by the joint degree distribution $P(k, k')$, i.e., as the
 probability of a link connecting two nodes of degree $k$ and $k'$. Another way to
 express this is by giving the conditional probability that an arbitrary neighbor
of a node of degree $k$ has degree $k'$
\begin{equation}
P(k \| k^{\prime}) = { {P(k,k^{\prime})} \over {P(k)}}
\end{equation}
This distribution gives a
very detailed description of node degree correlations but, for fat tailed degree
distributions as in scale-free networks, it is diffcult to evaluate experimentally,
because of the poor statistics. A measure with better statistics is to computing
the mean degree of the neighbors of nodes with a given degree, given by
\begin{equation}
k_{nn}(k) = \sum_{k^\prime}{k^{\prime} P( k^{\prime} | k)}
\label{vertDegree}
\end{equation}
 In the simulation we use Eq. (\ref{vertDegree}) for calulating vertex degree distribution.
 The vertex degree distribution can be calculated for the self degrees and neighbour degrees separately. For a random or small worlds network, the vertex degree distribution is a Poisson distribution in a large N limit with a peak at a particular degree bein most prominent. In contrast, for the scale free network of Barabasi and Albert, the degree distribution follows a power law for large k. In the network model that we consider we can get both types of degree distribution depending on parameter regimes. This is demonstrated in Figs 2(a) and 2(b). the Fig 2(a) shows a typical small worlds behaviour whereas 2(b) shows a power law like decay in large k region. Thus we can model both kinds of phenomena with the same model. As one changes the coupling, we get a transition from small worlds phase to a scale free phase. Further investigations of this transition are in progress.

 {\sc Clustering Coefficient}
 \vskip 0.1in
 A characteristic of the Erdos-Renyi model is that the local structure of the
 network near a node is a tree. More precisely, the probability of loops involving
 a small number of nodes tends to 0 in the large network size limit. This is in
 marked contrast with the profusion of short loops which shows up in many real-
 world networks. One way to characterise the presence of such loops is through
 the clustering coefficient.
 Two different clustering coefficients are frequently used. we can define a coefficient 
 \begin{eqnarray}
 C &= {3 N_{t} \over N_{3}} \nonumber \\
 N_{t} &= {\sum_{i,j,k}{a_{ij}a_{jk}a_{ik}}} \nonumber \\
 N_{3} &= \sum_{i,j,k}{a_{ij}a_{jk}} \nonumber \\
 \label{triangle}
 \end{eqnarray}
 
 This coefficient gives the ratio of number of {\it triangles} i.e. the triples which are linked to each other to number of connected triples (i.e. three nodes connected to each other directly or indirectly). If we denote by $l_{i}$ the number of links between neighbours of node i and $k_{i}$ is the number of neighbours of node i then we can also write clustering coefficient as
\begin{eqnarray}
C_{i} &= {2 l_{i} \over {k_i (k_i -1)}} \nonumber \\
C &= {1 \over N} \sum_{i} {C_{i}}
\label{total}
\end{eqnarray}

The difference between the two definitions is that Eq. (\ref{triangle}) gives the same weight to each triangle in the network, while Eq. (\ref {total}) gives the same weight to each node, resulting in different values because nodes of higher degree are possibly involved in a larger number of triangles than nodes of smaller degree. In the simulation result we calculate the coefficient given by Eq. (\ref{total}).

{\sc Centrality}
 \vskip 0.1in
  Greater the number of paths in which a node or link takes part, the greater the importance of this
 node or link for the network. Assuming that the interactions follow the shortest
 paths between two nodes, it is possible to quantify the importance of a node in this sense by its {\it betweenness centrality}
 
 \begin{equation}
 B_{i} = {\sum_{jk}{ \sigma(j,i,k) \over \sigma(j,k)}}
 \end{equation}

 where $\sigma(j,i,k)$ is the number of shortest paths between nodes j and k which pass through node $i$ and $\sigma(j,k)$ is total number of paths between nodes j and k. The sum is over all distinct pairs (j,k). Other centrality measures are given by {\it closeness centrality D , network centrality N and stress centrality S}
 \begin{eqnarray}
 D_{i}& = \sum_{j} { {1 \over {d _{ij}}}}   \nonumber \\
 N &= ({max_{j} d_{ij}})^{-1} \nonumber \\
 S &= \sum_{jk}{\sigma(j,i,k)}
 \end{eqnarray}
 
 where $d_{ij}$ =  shortest distance between i,j
 
 \subsection{Geometric characterisers}
 
 {\sc Geodesic Lengths}
 
 In the general case, two nodes of a complex network are not adjacent. In fact, most of the networks of interest are sparse, in the sense that only a small fraction of all possible links are present. Nevertheless, two non-adjacent nodes $i$ and $j$ can be connected through a sequence of m links (i, k1), (k1, k2), . . . , (km-1, j); such set of links is called a path between i and j, and m is the length of the path. We say that two nodes are connected if there is at least one path connecting them. 

A geodesic path (or shortest path)
between nodes i and j, is one of the paths connecting these nodes with minimum length. The length of the geodesic paths is the geodesic distance $d_{ij}$ between nodes i and j. There can be more than one geodesic paths between two nodes. We take the mean geodesic distance as a characteriser of the network. The definition has a problem if the network contains unconnected nodes therefore this measurement is only performed after cleanup operation. The Dijkstra algorithm is used for calculating the shortest path on the network.
 
  {\sc Network Efficiency}
 
 Network efficiency measure the ease of communication on a network. This can be defined as
 
 \begin{equation}
 \xi = {1 \over {N(N-1)}} \sum_{i \neq j}{{d_{ij}}^{-1}}
 \label{efficiency}
 \end{equation}
 
 This quantity is also useful in characterizing robustness of network to attack (see below)
 
 \subsection{Entropic characterisers}
  
   {\sc Growth rate and Entropy}
  
   The entropy of a network can be calculated by using the following result based on Frobenius - Perron theory  which we state without proof. for proof see \cite{lindandmarcus}
   
   {\bf Theorem} :

   {\it If G is an irreducible graph then $ h(X_{G}) = log \lambda_{A(G)}$ where $ h(X_{G})$ is the entropy of the shift dynamical system $X_{G}$ associated with the graph $\lambda_{A(G)}$ is the Perron eigenvalue of the graph( i.e. in this case the largest real eigenvalue of the adjacency matrix).
   
   }
   
   In general, however the above result provides an upper bound to entropy and we calculate this by diagonalising the adjacency matrix and finding the largest eigen value. Since our network is an evolving hypergraph , we compute this quantity at every time step and plot it with time in Fig 3. We find that the entropic bound saturates to certain fixed value (see Table 1) for a set of parameters and grows without bound for another set of parameters. This means that the network attains certain optimal size for a set of parameter values and has unrestrictive growth for another set.
   Continuously varying the parameters we have a transition from restrictive to unrestricted growth. 
   
   {\sc Generalised Entropy}
  
  Our network starts with nodes having a degree $<=$2. the degrees are added by splitting and by merger and removed by nodes dying.
  In the parameter regime where we get scale free behaviour the degree distributions can be fitted by
  \begin{eqnarray}
  P(\geq k) &= e_{q_c}^{(k -2)/K} \nonumber \\
   e_{q_c}^{x}&= [1+(1-q_c)x]^{1 \over {1 -q_c}} \nonumber \\
   for \nonumber \\
   1+(1-q_c)x & > 0 
   \label{dist}
  \end{eqnarray}
  
  where K is related to the the value $q_{c}$ which best fits the data. 
  This quantity is determined as follows: we take the logarithm of the probability  for a series of different values of $q_c$ ranging from -3 to 3 in steps of 0.1. The values obtained from the q -exponential form are then fitted to data obtained from the model by a linear fit. The value of $q$ which gives the best fit is determined as value $q_c$
      
   This form is taken from the formulation of a non-extensive genrealised entropic function introduced by Tsallis {\cite {tsallis}}
  
 The  entropic function
   \begin{eqnarray}
   S_q \equiv { {1 - \int_{2}^{\infty} {dk [p(k)]^{q}}} \over {1 -q}}
   \end{eqnarray}
  extremising this  as described in {\cite {tsallis}}  gives
  \begin{eqnarray}
  P(\geq K) = [1 -(1-q) \beta (k-2)]^{ (2-q) \over (1 - q)}
  \label{nent}
  \end{eqnarray}
  
 this equation is identical to Eq.(\ref{dist}) if we assume $ q_c = 1 / (2-q)$ and $ K = 1/ ( 2- q) \beta $. Thus this non-extensive entropic function reproduces the behaviour of probability distribution.  Therefore, it is probable that the correct entropic form for these models is non-extensive Tsallis entropy rather than by Shannon - Boltzmann -Gibbs entropy.
 The consequences of this conjecture require further investigations.
 
 \section {Response to Attack}
 
 One of the key features of any network is its ability to withstand various kinds of attacks. An attack or damage to a network typically hampers its performance which leads to a decrease in its efficiency of communication or transfer of signals along various available paths in the network. Self organised networks have various responses depending on the types of attack. Needless to say, this characteristic is of utmost importance for biological contexts where the response to an external attack could determine the survival, recovery or viability of a system. A number of attack strategies are usually considered while evaluating the response. These include targeted attacks (where a particular node or connected segment is attacked based on some of its properties like vertex degree or length), multiple attacks (where random nodes are targeted), isolated attack (where a single node is targeted) etc. In the current simulation we consider isolated and multiple attacks. The measure of response to attack is given by network efficiency as defined in Eq. (\ref{efficiency}). The response of the network to these attacks is shown in the table where the average change in network efficiency with the number of nodes attacked is shown. This quantity is computed by choosing many nodes either one at a time  or in a bunch
 and averaging over the decrease in network efficiency. As can be seen from the table, the network is more susceptible to multiple random attacks rather than to isolated single ones. This is a further manifestation of the fact that the spatio-temporal patterns found in the network are robust with respect to external perturbation such that removal of an element from a local pattern does not impair the network as much as removal of multiple randomly placed elements does. This is also indicative of the fact that there are more than one signalling paths available in any local neighbourhood and unless one blocks a large number of them the efficiency is not significantly impaired.
 
 \section{Conclusion}
 
 Therefore, in this paper we have proposed a generic model reproducing qualitative behaviour of collective biological systems. The network incorporates a new feature of mobility which has been hitherto missing from such models. This property is shown to be essential in reproducing behaviour in presence of diffusion which many contexts show. The network goes beyond the assumption of boolean elements. It is shown that some of the shortcomings found in boolean networks can be overcome by making the state of an element a continous variable. Prominient among such enhancements is an ability to model highly heterogenous and diverse environment where the internal state determines and is determined by its surroundings, as in a cellular interaction mechanism. At the same time the network allows us to have a number of states with switching attractors. Thus this model mimics the ``robust yet fragile" nature of biological systems.
 
  This explicit coupling between intra and inter dynamics also gives an insight into several topological features of such networks. In particular, we have demonstrated that starting from a particularly simple initial configuration and without explicitly assuming any arbitrarily constructed rule for adding nodes, we can obtain both scale free nature as well as networks containing important hubs. The latter being reminiscent of small world networks. We present evidence for a transition between two regimes upon variation of parameters.
  
   We also show that interesting cooperative phenomena like community structures and synchronisation emerge in these types of networks. This is clearly seen in Fig. 4 showing the spatial profile of internal variables. In this figure the regions of closely linked variations are distinctly separate from other regions.
   
   To investigate robustness of such a network to external perturbations we present a study of two different types of attack on the system. It is found that the multiple random attack is more effective than an isolated one. In current networks the reponse to the attack is reflected by a decrease in its efficiency. 
   
   The growth and size  of the network follows an entropic law which shows signatures of non-extensivity. This leads to the speculation that Tsallis statistics is probably more relevant to biological networks.

    Several of these interesting features need further exploration and investigation which are currently under progress. However, in the present paper we only report that many qualitative aspects of the diverse behaviour of cooperation in biological systems can be modeled by a single generic network model. Various aspects of this model can give us an insight into important questions like quorum sensing in bacteria, synchronisation of biological elements, emergence of heterogeniety and morphogeneis, tissue and cell growth, tumour growth, protein interaction pathways and biochemial networks.
 
    For the network theory, this paper presents a network which shows number of different, well known network characteristics in a single model. Therefore, this effort may help in unifying some of different pieces of knowledge which we hope would be useful in understanding general networks better.
    
    Both authors contributed equally to the work. Supplementary materials like programs, data set and additional figures are freely available from the authors upon request.
    \newpage
    \section{Figure Captions}
    \begin{enumerate}
    {\item {{\bf Figure 1:}  Figure 1 describes the flow of the network evolution. The blocks in the figure describe how various stages of dynamics are handled and how they depend on each other.}}

    {\item {{\bf Figure 2:}  Figure 2 describes the typical vertex degree distributions obtained. The figures 2(a) and 2(b) are generated by initially setting up a golden mean network with two different kinds of node,namely the Mutual activation switch Eq.(\ref{MASwitch}) and  the activator - inhibitor oscillator Eq.(\ref{AIOscillator}). The network was evolved with value of R = 0.35, and the same values of node parameters as in \cite{tyson}. The value of the coupling strength $\epsilon$ was 0.66 for 2(a) and 0.033 for 2(b). Each node was evolve with respect to internal dynamics by using a fourth order unge-Kutta method. The step size for the switch was 0.1 and for the oscillator 0.001.}}

     {\item {{\bf Figure 3:} Describes the entropic bound for growth as described in the paper. Fig. 3(a) describes the behaviour where the growth of the network saturates after some time and Fig. 3(b) where it grows without bounds.}}

    {\item {{\bf Figure 4:} shows the patterns found in internal profile of the network at various times (a)-(e). The presence of correlated patterns can be easily seen from this figure.}}
    
    \end{enumerate}



\newpage

\begin{figure}
\includegraphics[]{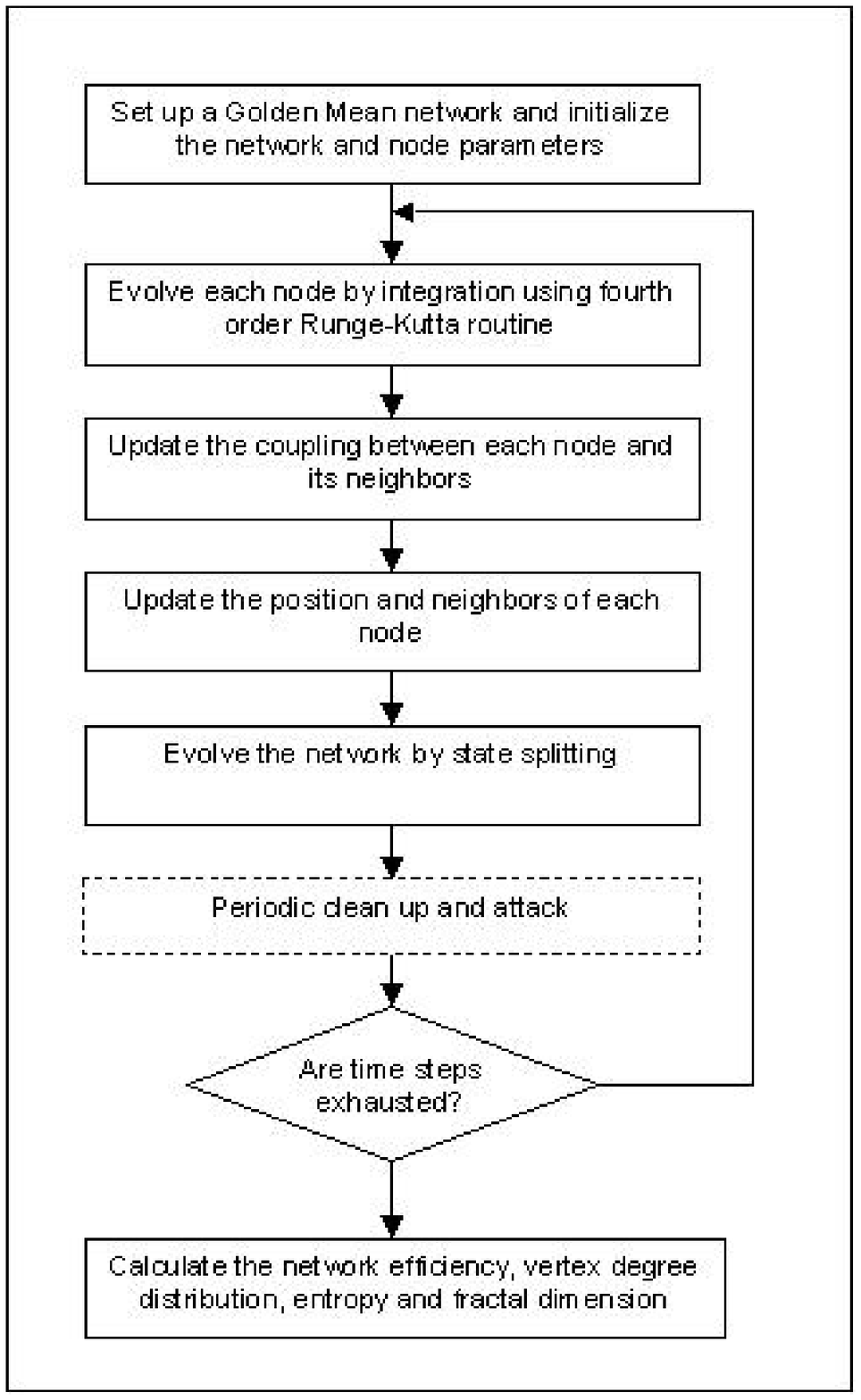}
\caption{}
\end{figure}

\begin{figure}[htbp]
  \centering
    
      \subfigure[]{\includegraphics[scale=0.8,angle=270]{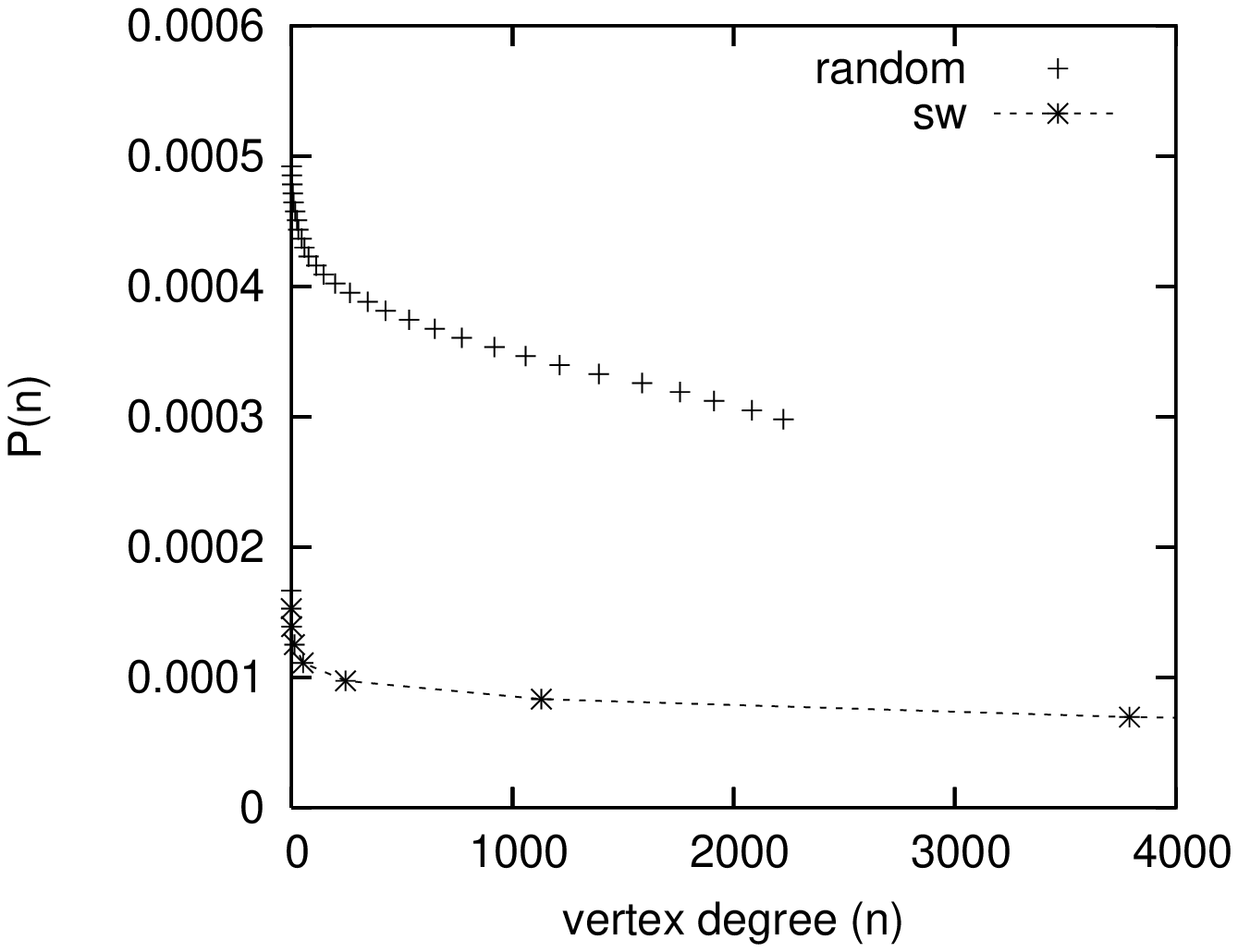}}
      \subfigure[]{\includegraphics[scale=0.8,angle=270]{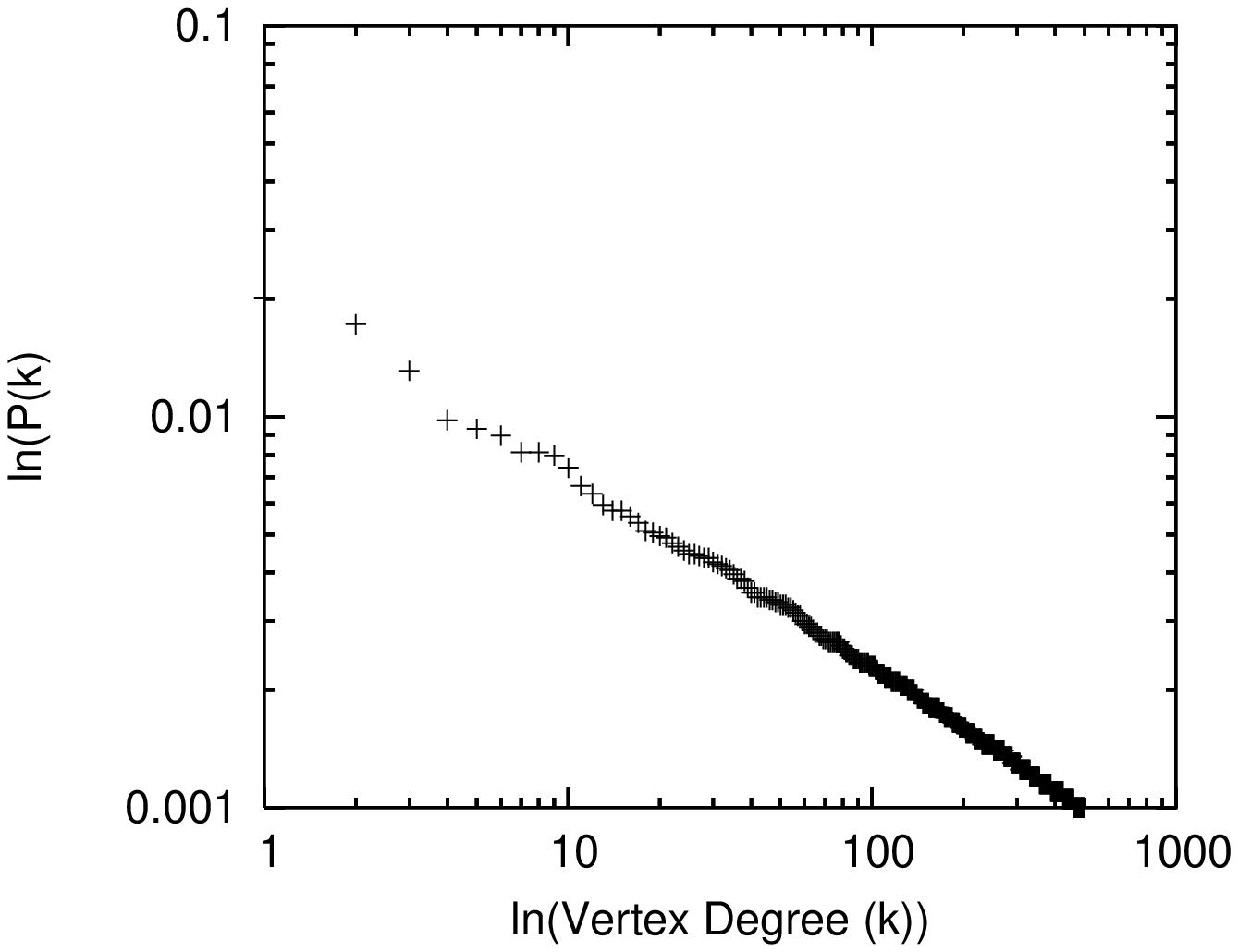}}
    
    \caption{}    
\end{figure}

\begin{figure}[htbp]
  \centering
    
      \subfigure[]{\includegraphics[scale=0.8,angle=270]{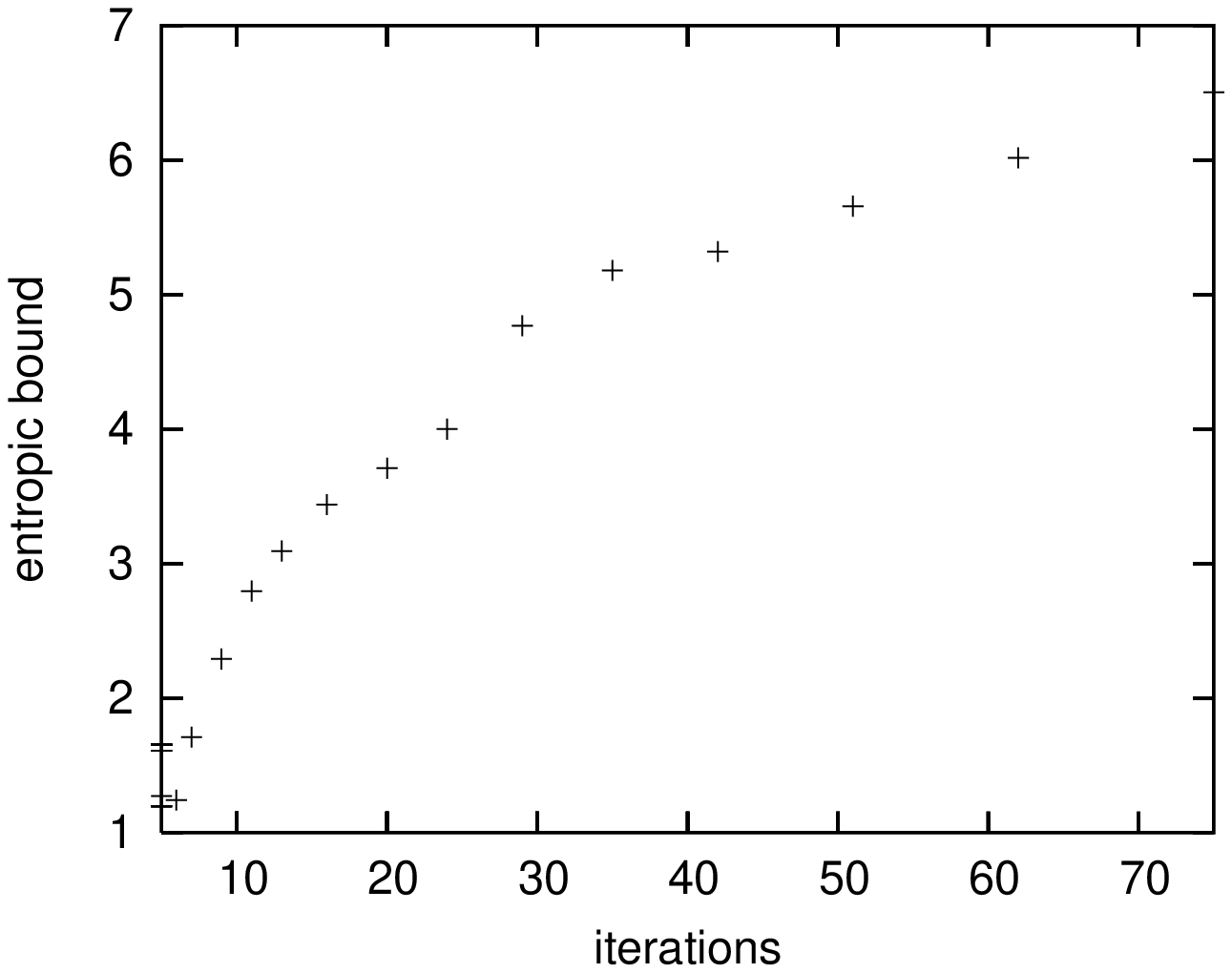}}
      \subfigure[]{\includegraphics[scale=0.8,angle=270]{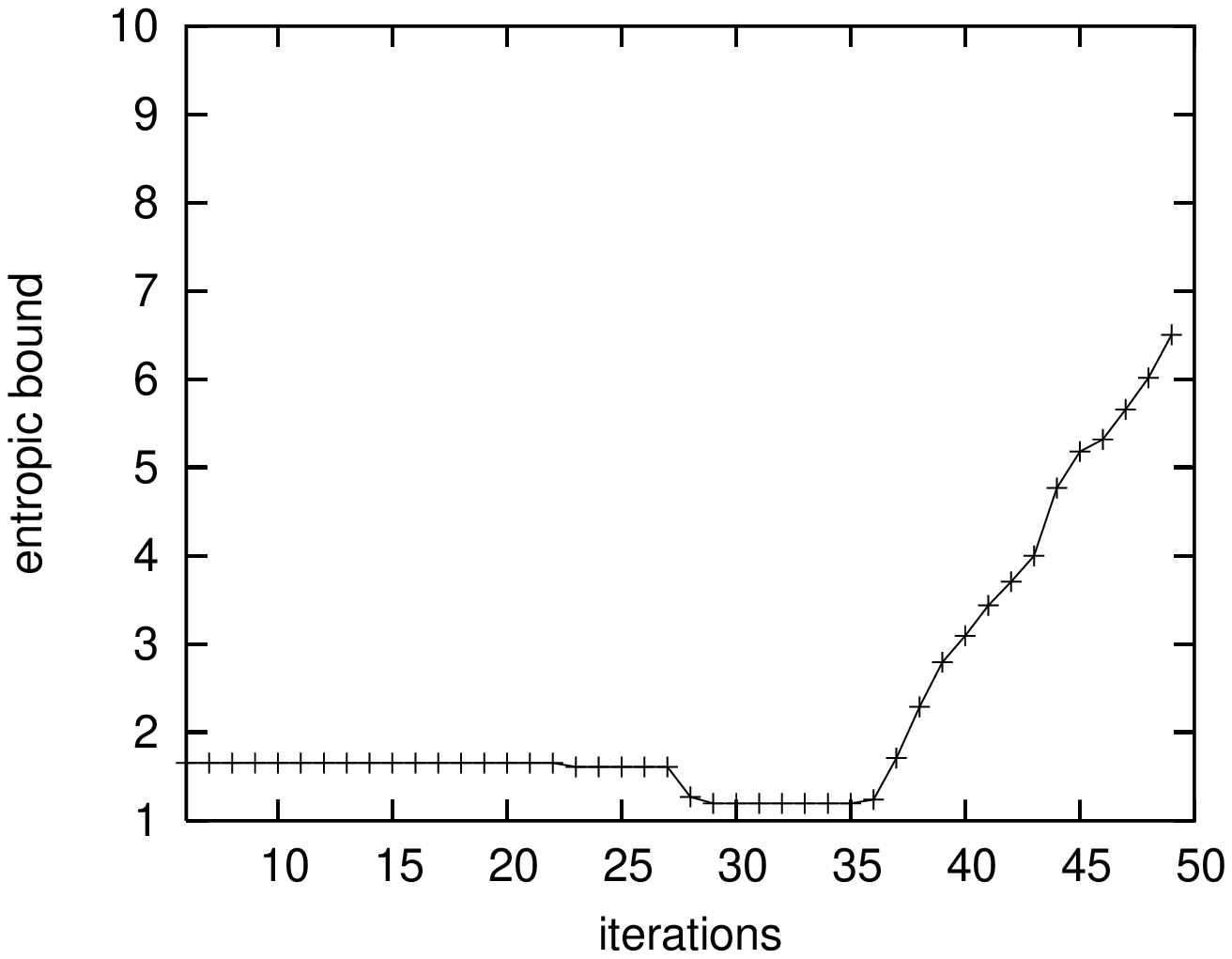}}
    
    \caption{}    
\end{figure}

\begin{figure}
\includegraphics[angle=270]{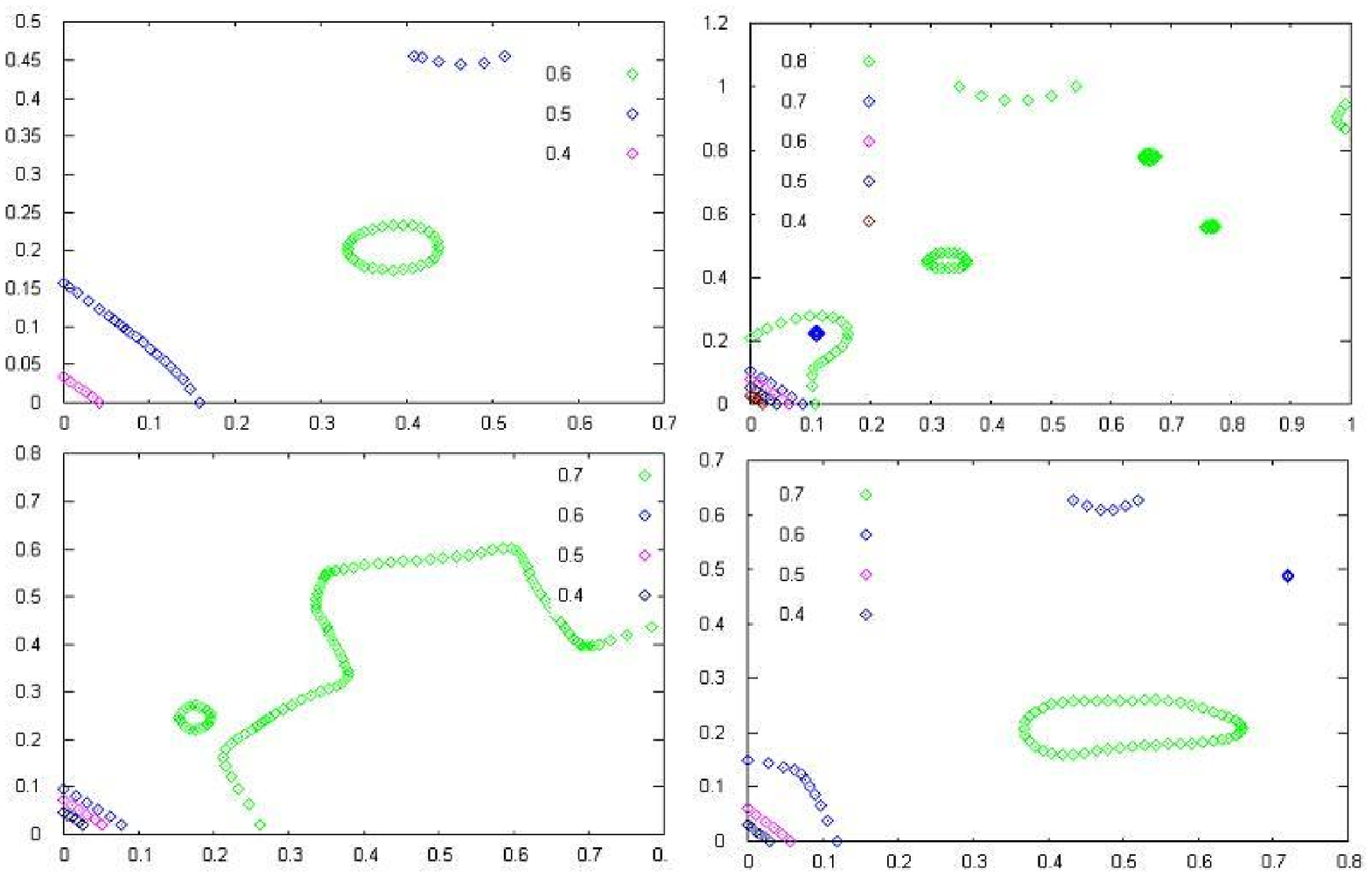}
\caption{}
\end{figure}

\end{document}